# On the origin of tail states and $V_{OC}$ losses in Cu(In,Ga)Se$_2$

*Omar Ramírez*[*], *Jiro Nishinaga, Felix Dingwell, Taowen Wang, Aubin Prot, Max Hilaire Wolter, Vibha Ranjan and Susanne Siebentritt*

O. Ramírez, F. Dingwell, T. Wang, A. Prot, M. H. Wolter, V. Ranjan and Prof. Dr. S. Siebentritt
Department of Physics and Materials Science, University of Luxembourg, 41 rue du Brill, L-4422, Belvaux, Luxembourg
E-mail: omar.ramirez@uni.lu

Dr. J. Nishinaga
Research Institute for Energy Conservation, National Institute of Advanced Industrial Science and Technology (AIST), Koriyama, Fukushima, 903-0298, Japan.

The detrimental effect of tail states on the radiative and non-radiative voltage loss has been demonstrated to be a limiting factor for the open circuit voltage in Cu(In,Ga)Se$_2$ solar cells. A strategy that has proven effective in reducing tail states is the addition of alkali metals, the effect of which has been associated with the passivation of charged defects at grain boundaries. Herein, tail states in Cu(In,Ga)Se$_2$ are revisited by studying the effect of compositional variations and alkali incorporation into single crystals. The results demonstrate that alkalis decrease the density of tail states despite the absence of grain boundaries, suggesting that there is more to alkalis than just grain boundary effects. Moreover, an increase in doping as a result of alkali incorporation is shown to contribute to the reduced tail states, which are demonstrated to arise largely from electrostatic potential fluctuations and to be determined by grain interior properties. By analyzing the voltage loss in high-efficiency polycrystalline and single crystalline devices, this work presents a model that explains the entirety of the voltage loss in Cu(In,Ga)Se$_2$ based on the combined effect of doping on tail states and $V_{OC}$.

# 1. Introduction

Generation and recombination processes are the basis on which our understanding of solar cells and fundamental models like the Shockley-Queisser limit finds its origins.[1] Understanding how carriers recombine and finding ways to reduce detrimental non-radiative recombination channels, have been part of the roadmap that has led to the improvement in efficiency in all solar cell technologies. Among the possible sources of non-radiative recombination, available energy states that extend from the bands into the band gap, known as tail states, have proven to be of great importance in describing the recombination activity in different kinds of semiconductor materials such as chalcopyrites[2-3], amorphous silicon[4], perovskites[5-6] and organic.[7] More importantly, the detrimental effect of tail states is not only limited to non-radiative recombination but also affects the radiative open-circuit voltage, which is reduced as a result of the increased density of states available for absorption below the band gap.[2, 8-9] Tail states are associated with the thermal, structural and electrostatic disorder present in the material and can describe how far the material is from a perfect crystal.[10] Even though it is not possible to measure the density of tail states in a direct way, the relation between the absorption coefficient $\alpha$ and the joint density of states, enables absorption-based techniques like photoluminescence (PL)[11] to access this information through the determination of the Urbach energy ($E_U$)[12], as depicted in Figure 1(a). In other words, the higher the Urbach energy, the higher the density of tail states present in the material.

Recently, an empirical linear relationship between Urbach energy and open-circuit voltage losses has been observed in different materials such as perovskites, kesterites, PbS quantum dots and chalcopyrites.[3, 6, 13-16] Even though it is often assumed that Urbach energies below $k_B T$ do not play a role, it has been demonstrated that they do in fact matter, and that each additional meV in Urbach energy causes an additional 20 mV voltage loss for Urbach energies in the range of 10 to 20 meV.[2, 6] In the specific case of Cu(In,Ga)Se$_2$, for example, it has been proposed that the open-circuit voltage loss can be entirely attributed to tail states.[2, 17] The fact that a lower $E_U$ can be associated with reduced voltage losses, has prompted calls for the Urbach energy to be used as a parameter to describe the absorber's quality.[18] Since optical techniques can be used directly at the absorber alone, without finishing the device, the Urbach energy can be used for the fast-screening of materials, similarly to what has been proposed in literature for the optical diode ideality factor[19].

The recent efficiency improvements in Cu(In,Ga)Se$_2$ solar cells, including the current record of 23.35%[20], are related to the development of alkali metal postdeposition treatments (PDTs), which have been found to affect both the surface and bulk properties of the absorber.[17] The most striking effect of PDTs is, perhaps, the one related to the improvement in open-circuit voltage ($V_{oc}$), which has been associated to the passivation of charged defects at grain boundaries.[17, 21] The idea behind this argument is based on the observation of alkali segregation at grain boundaries by atom probe tomography (APT)[17, 22], as well as changes in band-bending measured by Kelvin probe force microscopy[17, 23], phenomena that lead to the reduction of tail states and thereby to the aforementioned $V_{oc}$ improvement. Evidence on the reduction of tail states caused by alkali incorporation can be found in literature in the form of a reduced Urbach energy after the treatments.[2, 17, 24] However, one of the consequences of the accumulation of heavy alkali elements at grain boundaries is the increase of the sodium concentration in the grain interiors.[17] An increase from ~15 ppm to ~40 ppm after a rubidium fluoride (RbF) postdeposition treatment, for example, was detected by means of APT.[22] This phenomenon, although reported, has not been taken into account when explaining the beneficial effect of PDTs on the $V_{oc}$. Furthermore, the fact that the effectiveness of PDTs has been demonstrated in single crystals, where there are no grain boundaries, provides further evidence that there is more to alkali PDTs than just grain boundary effects[25-26].

In this contribution, we revisit the tail states in Cu(In,Ga)Se$_2$ by a photoluminescence study of the Urbach energy. In the first part, we analyze how compositional variations affect the Urbach energy of alkali-free single crystals, finding that increasing the Ga/(Ga+In) and reducing the Cu/(Ga+In) ratio result in an increased Urbach energy. After that, a temperature and intensity-dependent photoluminescence study is carried out in order to investigate the higher Urbach energy measured in Cu(In,Ga)Se$_2$ when compared to pure CuInSe$_2$, finding that Cu(In,Ga)Se$_2$ suffers from more severe electrostatic potential fluctuations. Then, the effects of sodium fluoride (NaF) and potassium fluoride (KF) on the Urbach energy and quasi-Fermi level splitting (qFls) of single crystals are studied. It is revealed that NaF, and to a lesser extend KF, can effectively decrease the Urbach energy of single crystals in a similar way as they do in polycrystalline absorbers. Furthermore, an increase in doping as a result of alkali incorporation is shown to be associated to the reduced tail states after the treatments. This is explained based on the fact that as a consequence of the increased doping concentration, a reduced degree of compensation results in a lower density of charged defects or more screening of charges, which in turn reduces the magnitude of the electrostatic potential fluctuations and thus, also the tail states. Finally, a model explaining the entirety of the voltage loss in Cu(In,Ga)Se$_2$ based on the combined effects of doping on tail states and the $V_{OC}$ is presented. Our analysis of the $V_{OC}$ losses of high-efficiency polycrystalline and single crystalline devices shows, that tail states in Cu(In,Ga)Se$_2$ are not mainly determined by grain boundaries and that changes in the sodium concentration at the grain interiors after PDTs is actually the main driver of the V$_{OC}$ improvement. The aim of this work is to provide a thorough investigation on what can affect the density of tail states in Cu(In,Ga)Se$_2$ and what can be done in order to reduce the V$_{OC}$ loss.

## 2. Tail states and Urbach Energy

The density of states within the band gap of an ideal semiconductor is zero. In a real semiconductor however, a density of states that decays exponentially towards the gap exists. Such energy states are known as tail states or Urbach tails, and they were first described by Franz Urbach while studying the absorption edge of silver bromide crystals.[12] The characteristic decay energy of the absorption edge is the so-called Urbach energy $E_U$, which relates to the thermal and structural disorder in the crystal.[10, 27] The thermal component describes the occupation of phonon states (vibrational disorder), while the structural disorder adds a temperature independent contribution caused by the structural deviation from the ideal crystal, such as the presence of crystallographic defects.

In the case of Cu(In,Ga)Se$_2$, such structural disorder can be the result of, for example: i) grain boundaries[28] ii) variations of the Ga-Se and In-Se bond length [29], iii) inhomogeneities in the elemental composition that would cause local fluctuations of the band gap (as it depends on the Ga content)[30] and iv) random distributions of charged defects that result in fluctuations of the electrostatic potential.[31-33] Thus, to understand how compositional variations affect the density of tail states, the following subsection deals with how the Urbach energy of Cu(In,Ga)Se$_2$ is affected by changes in the copper and gallium content. Furthermore, the analyzed samples consist of single crystals in order to exclude the effects of grain boundaries. The evidence for the single-crystal character of the studied samples can be found elsewhere [26, 34].

### 2.1. Tails in alkali-free Cu(In,Ga)Se$_2$ single crystals

Tail states arise from the thermal and structural disorder present in the crystal. Following this idea, it would be expected that differences in the elemental composition of Cu(In,Ga)Se$_2$ absorbers would have an influence on the tail states and therefore, on the measured Urbach energy. Replacing copper with silver, for example, has been reported to effectively decrease the chalcopyrite disorder since lower Urbach energies were measured in Ag-containing samples when compared to pure Cu(In,Ga)Se$_2$ with the same amount of gallium[35].

Nevertheless, the influence of the Cu content is, perhaps, the most well-known parameter that has a strong influence on the Urbach energy. It has been demonstrated in pure CuInSe$_2$ that the Urbach energy increases with decreasing Cu content, and that it remains essentially constant from the stoichiometric point towards Cu-rich compositions. [28] The reason behind this, lies in the fact that Cu-poor material has a higher density of both donor and acceptor defects which results in a higher degree of compensation. [32, 36-37] As an example, the absorption coefficient $\alpha$ of different Cu(In,Ga)Se$_2$ single crystals with Ga/(Ga+In)~0.4 and varying copper content is displayed in **Figure 1**(a). As can be seen, the absorption onset becomes steeper with the increase in copper content, resulting in a lower Urbach energy as determined from fitting $\alpha$ with $e^{E/E_U}$ (details on the determination of $E_U$ from PL can be found in section S1 of the supporting information and references [2, 11]). As a way to quantify the effect of the copper content on the Urbach energy, a comparison between CuInSe$_2$ (data taken from Figure 4 in reference [28]) and CuIn$_{0.6}$Ga$_{0.4}$Se$_2$ single crystals was carried out and is shown in Figure 1(b). Interestingly, the effect of copper deficiency appears to be similar in both compositions, where an increase of around 2 meV in Urbach energy for every 0.1 decrease in copper content is measured.

Since the Urbach energy is closely related to disorder, it could be expected that alloy disorder arising from In/Ga intermixing would have an influence on the Urbach energy in a similar way as the Cu content does. In order to investigate this influence, which can be glimpsed already from Figure 1(b), the Urbach energy of several alkali-free single crystals with varying Ga/(Ga+In) was also investigated. Since the Cu content has a strong effect on $E_U$ as previously discussed, the Cu/(Ga+In) ratio was limited to values close to the stoichiometric point (from 0.95 to 1.07). As some of the samples were grown with a Ga-gradient, it was decided to use the maximum energy of the photoluminescence as a proxy for the band gap ($Eg_{PL}$) and thus as an indicator of the Ga content instead of the Ga/(Ga+In) measured by EDX, since the luminescence measured in samples grown with a Ga-gradient comes from the area with the lowest band gap. [38] With the extracted Urbach energies, the interpolated color map of **Figure 2**(a) was constructed. Even though only a narrow range of Cu composition is studied, its effect on $E_U$ can still be observed (horizontal). For a fixed Cu/(Ga+In) composition, the figure shows also the effect of the gallium content on $E_U$ through an increase of the Urbach energy towards higher gallium concentrations (vertical). The samples' composition, $E_U$ and $Eg_{PL}$ values can be found in Table S1 of the supporting information.

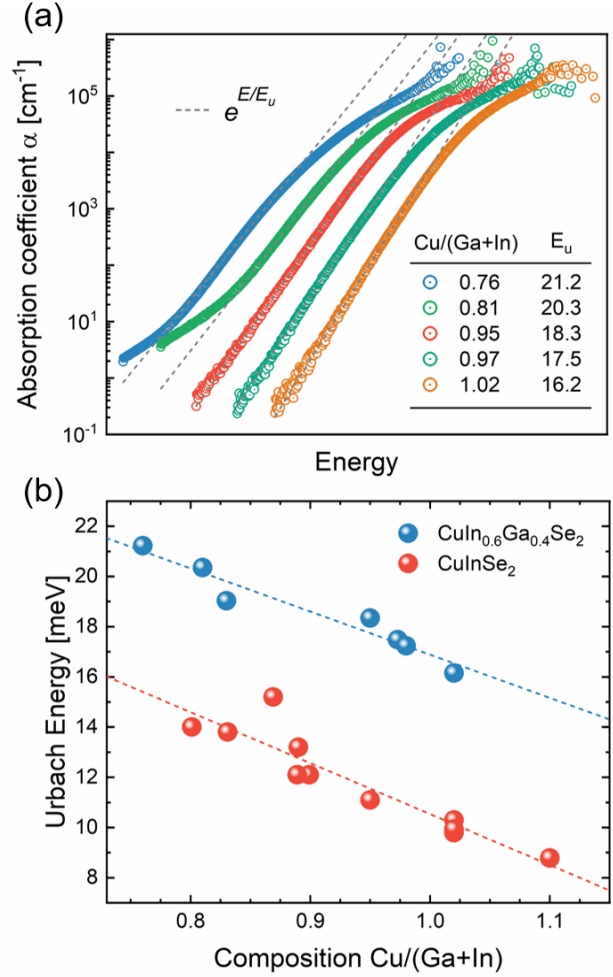

**Figure 1.** Absorption coefficient derived from room temperature photoluminescence measurements of a copper series of Cu(In,Ga)Se$_2$ single crystals with ~40% gallium. Dashed lines are fits from which the Urbach energy is extracted. Data are shifted in energy for visualization purposes (a). Urbach energy comparison between CuInSe$_2$[28] and Cu(In,Ga)Se$_2$ as a function of the copper content. Dotted lines are guides to the eye (b).

Even though it is reasonable to assume that the increase in $E_U$ due to the Ga content is related to alloy disorder, it could also be related to the increase in band gap that goes along with it. An increasing trend of the Urbach energy with the band gap has been experimentally observed in other materials like kesterites[13-14] and perovskites.[5] Figure 2(b) gathers the Urbach energy as a function of the band gap of some of the aforementioned materials and references. Attention should be paid to the trends rather than to the magnitude of $E_U$, as the methods used to obtain the Urbach energy are different (photoluminescence, photothermal deflection spectroscopy and transient photocapacitance spectroscpopy) and so is their sensitivity to measure low absorption coefficients. [11] In the case that the change of $E_U$ in Cu(In,Ga)Se$_2$ would be caused only by alloy disorder, the Urbach energy of pure CuGaSe$_2$ would be expected to be lower than in an alloy with 50% indium, for example. By extrapolating the value of $E_U$ in Figure 2(b) to the band gap of CuGaSe$_2$ (~1.7 eV)[39], an Urbach energy of around 26 meV would be expected (gray star in Figure 2(b)), which is surprisingly close to the reported values for polycrystalline CuGaSe$_2$ absorbers measured by photodeflection spectroscopy (25-35meV)[40] and ellipsometry (33.8meV).[15] Even though the data for Cu(In,Ga)Se$_2$ are based on single crystals and the reported values are from polycrystalline absorbers, it has been shown that the Urbach energy is higher in the latter but not so much when the composition is close to the stoichiometric point[28], which is the case here. For a final decision on the role of alloy disorder vs. band gap, in a future study Ga/(Ga+In) ratios above 50% grown by the same process must be studied. But the available data points out that the increasing trend in $E_U$ observed with the increase of gallium content in Cu(In,Ga)Se$_2$ seems to be an intrinsic property of the continuously increasing band gap rather than caused by alloy disorder. The discussion on the dependency of the Urbach energy on the band gap and the origin of the larger Urbach energies measured in samples with higher Ga contents will be addressed in the following subsections.

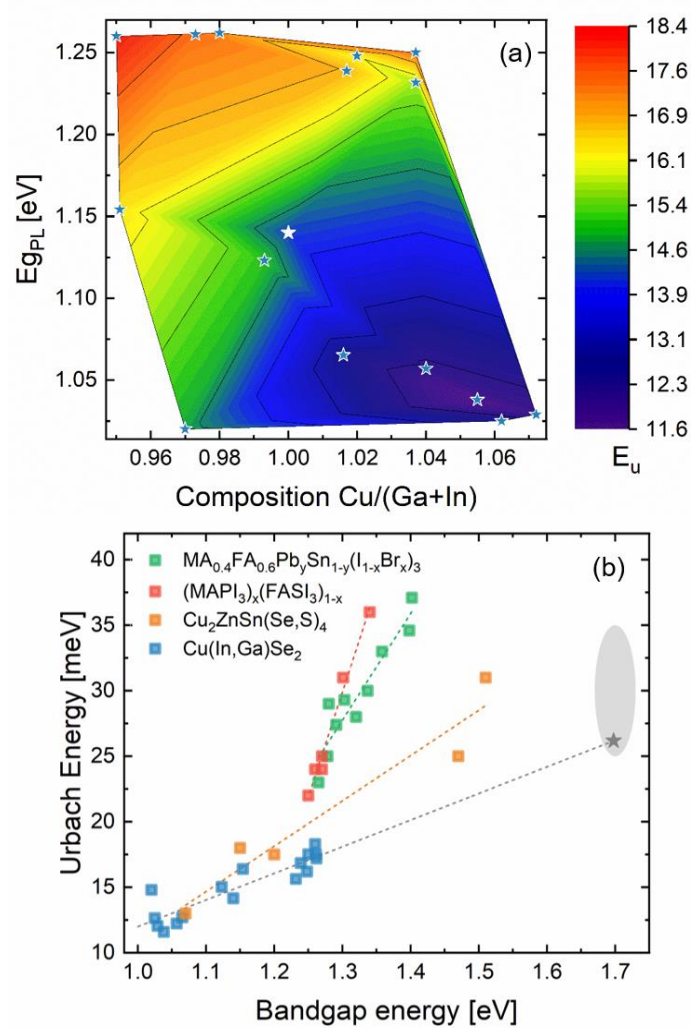

**Figure 2.** Interpolated Urbach energy map as a function of the Cu-content and optical band gap based on alkali-free Cu(In,Ga)Se$_2$ single crystals grown by metalorganic vapor phase epitaxy (blue stars) and molecular beam epitaxy (white star) (a). Urbach energy as a function of the band gap in some Sn-based perovskites[5], kesterites[13] and chalcopyrites (b). The data for Cu(In,Ga)Se$_2$ is the same as in (a). The gray star is the extrapolated Urbach energy for CuGaSe$_2$ and the shadowed area denotes the experimental values[40] reported by Meeder et al.

The results discussed so far, especially the Urbach energy map in Figure 2(a), would suggest CuInSe$_2$ with copper contents close-to-stoichiometric or even Cu-rich as a great solar cell absorber in order to minimize the V$_{OC}$ deficit due to their lower density of tail states. However, it is known that the quasi-Fermi level splitting of Cu-rich Cu(In,Ga)Se$_2$ is lower than its Cu-poor counterpart and that devices made of Cu-rich material suffer from a higher V$_{OC}$ loss.[41] Also, it has been reported that Cu-excess in CuInSe$_2$ has a detrimental effect on the surface and interfaces which limits the device performance.[41-43] Nevertheless, recent advances in CuInSe$_2$ have demonstrated efficiencies above 19% [44] with a V$_{OC}$ deficit with respect to the Shockley-Queisser limit (considering a back reflector) of only 157 mV ($E_g$ = 1.0 eV, $V_{OC}$ = 609 mV, $V_{OC}^{SQ}$ = 766 mV)[44-45], which is just 50 mV higher than the current Cu(In,Ga)Se$_2$ record cell ($E_g$ = 1.08 eV, $V_{OC}$ = 734 mV, $V_{OC}^{SQ}$ = 841 mV).[20, 45] An important step in the optimization of this cell was to increase the Cu content as close as possible to the stoichiometric point[44].

**2.2. Tail states and Potential Fluctuations**

At the beginning of section 2, possible causes of static disorder in Cu(In,Ga)Se$_2$ such as potential fluctuations arising from charged defects or compositional variations were introduced. Then, in the previous subsection, it was established that the density of tail states is larger in Cu(In,Ga)Se$_2$ than in pure CuInSe$_2$ when grown with the same Cu content. In the following, a causal relationship between potential fluctuations and tail states is discussed. In order to study potential fluctuations, a temperature-dependent and excitation intensity-dependent photoluminescence analysis is carried out in several Cu(In,Ga)Se$_2$ single crystals with varying gallium content.

*2.2.1. Potential fluctuations from temperature-dependent PL*

As explained by Larsen et al.[33] based on the seminal work of Levanyuk and Osipov[46], the temperature dependence of the energy or the photoluminescence maximum in compensated chalcopyrites with potential fluctuations is not monotonic with increasing temperature but has a redshift-blueshift behavior. At very low temperatures (10K), photogenerated carriers are hardly mobile and defect-related recombination occurs from local potential valleys arising from the Coulomb interaction of charged defects. However, when the temperature is increased, carriers start to gain thermal energy that allows them to overcome the potential barriers and recombine from valleys of even lower energy, explaining the initial redshift. As the temperature continues to increase, the mobile free carriers screen the charged defects and flatten the potential fluctuations, which explains the subsequent blueshift of the PL peak.[33] Thus, the temperature at which the redshift stops ($T_{min}$) can be associated with the magnitude of the potential fluctuations, as deeper fluctuations will cause a higher $T_{min}$.[32] Besides the blueshift expected from the flattening of the fluctuations, an additional blueshift can be expected as the recombination path changes from defect-related to band-to-band. **Figure 3** shows the normalized temperature-dependent photoluminescence of three close-to-stoichiometric alkali-free single crystals with different gallium contents ranging from CuInSe$_2$ to CuIn$_{0.56}$Ga$_{0.44}$Se$_2$. Details on the construction of the intensity maps and the photoluminescence spectra can be found in section S2 of the supporting information. As can be clearly seen from this figure, the position of the PL peak in all samples shows the initial redshift behavior previously described. In the CuInSe$_2$ sample, $T_{min}$ is visible at around 80 K, and a clear blueshift is already visible at 100 K. On the other hand, in the sample with the highest gallium content, the blueshift is still not visible at this same temperature. Thus, based on the apparent increase of $T_{min}$ with the gallium content, a preliminary conclusion can be drawn: Regardless of the nature of the fluctuations, their magnitude increases with the gallium content.

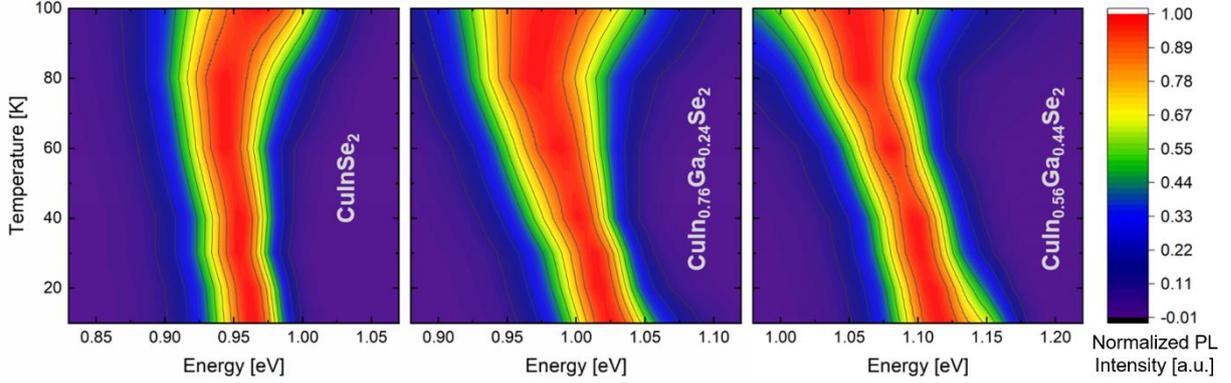

**Figure 3.** Temperature-dependent photoluminescence from 10 to 100 K of a CuInSe$_2$, CuIn$_{0.76}$Ga$_{0.24}$Se$_2$ and CuIn$_{0.56}$Ga$_{0.44}$Se$_2$ alkali-free single crystals with close-to-stoichiometric Cu contents. Each PL spectrum was normalized and the PL intensity is depicted by the colour code.

*2.2.2. Potential fluctuations from excitation intensity-dependent PL*

In order to corroborate the qualitative results from the temperature-dependent photoluminescence analysis, and provide more quantitative information about the magnitude of the potential fluctuations and their nature, an intensity-dependent photoluminescence study at low temperatures is needed. At low temperatures, the analysis of the shift of the PL peak energy position as a function of the logarithm of the excitation ($\beta$) provides information on the nature of the observed transition. Without fluctuations, most transitions do not shift with excitation. Nevertheless, the usual behavior of a donor-acceptor (DA) transition with increasing excitation is of a slight blueshift (~1-3 meV/decade of excitation in chalcopyrites[47-48]) due to the changes in the Coulomb attraction between the two charged defects that influences the photoluminescence peak position. However, in the case of radiative transitions perturbed by potential fluctuations, the flattening of the fluctuations with increasing excitation would also produce a blueshift, which is usually stronger than a DA transition without potential fluctuations.[47-48] The amount of the shift is another measure for the depth of the fluctuations[32].

In order to corroborate the results from the temperature-dependent analysis, the peak energy position as a function of excitation of the four samples listed in Table 1 was studied and is displayed in **Figure 4**(a). As can be seen, the magnitude of $\beta$ increases with the gallium content, confirming the conclusion already drawn by the temperature-dependent analysis: the magnitude of the potential fluctuations increases with the gallium content. Details of the peak position determination and the fitting of $\beta$ can be found in section S3 of the supporting information.

**Table 1.** Gallium content determined from EDX, optical band gap ($Eg_{PL}$) and Urbach energy ($E_U$) determined from room temperature PL. $\beta$ and $\gamma$ obtained from PL at 10 K of the four close-to-stoichiometric single crystals studied by intensity and temperature-dependent photoluminescence (Figure 4).

| Sample Number | Ga/(Ga+In) | $Eg_{PL}$ [eV] | $E_U$ [meV] | $\beta$ [meV/dec] | $\gamma$ [meV] |
|---|---|---|---|---|---|
| 285 | 0.01 | 1.03 | 12.6 | 2.5 | 20.9 |
| 288 | 0.24 | 1.12 | 15.0 | 7.6 | 22.2 |
| 275 | 0.38 | 1.24 | 16.8 | 12.5 | 23.4 |
| 282 | 0.44 | 1.26 | 17.2 | 13.1 | 22.6 |

For another approach to the magnitude of the potential fluctuations, a spectral fitting to the low energy wing of the PL can be performed. Depending on the depth of the fluctuations, the low energy wing can be spectrally fitted with either a Gaussian (for deep enough fluctuations) or with an exponential decay with decay constant γ.[32-33] It should be pointed out that $\gamma$ is not the Urbach energy $E_U$: $E_U$ is a property of the density of states while $\gamma$ is a property of the PL emission, which at low temperatures cannot be described by the generalized Planck's law due to the non-equilibrium between the different valleys arising from potential fluctuations. Furthermore, the fact that $\gamma$ is extracted from low-temperature measurements, indicates that the level of excitation is higher than at room temperature.

In order to obtain the magnitude of the potential fluctuations, we proceeded to fit the low energy wing of the same PL spectra used for the determination of $\beta$, finding that all of them can be well-described by an exponential decay as $I(E) \sim \exp\left(-(E - E_0)/\gamma\right)$, where $E_0$ is just an energy prefactor. Details on the fitting routine can be found in the supplementary information (section S3). As a result, we find that the magnitude of the potential fluctuations in these samples ranges from 21 to 23 meV at low excitation levels, which is in excellent agreement with values between 19 and 25 meV reported for $CuInSe_2$ and $CuGaSe_2$ epitaxial layers, respectively.[32] In agreement with our findings, it has been reported that in close-to-stoichiometric absorbers, the low energy wing of the PL can be better described by an exponential decay (independently of the Ga content)[32].

*2.2.3. Band gap versus electrostatic potential fluctuations*

By analyzing the dependency of $\gamma$ on excitation intensity, it is possible to obtain information regarding the nature of the potential fluctuations that causes the broadening of the PL emission, i.e., to distinguish whether the potential fluctuations are caused by charged defects (electrostatic) or by compositional variations (band gap). In the case of electrostatic potential fluctuations arising from the interaction of charged defects, it would be expected that $\gamma$ decreases with increasing excitation as an increase in photogenerated carriers will result in the screening of charged defects and, as a consequence, the flattening of the fluctuations.[49] Nonetheless, if the potential fluctuations are due to variations in the band gap, an increase in photogenerated carriers may flatten the fluctuations of one of the band edges, but it will not change the local band gap and will thus not flatten the band gap fluctuations. As can be seen in Figure 4(b), independently of the gallium content, the magnitude of the potential fluctuations decreases with increasing excitation, suggesting that the potential fluctuations observed in alkali-free $Cu(In,Ga)Se_2$ have an electrostatic nature. Interestingly, the decrease in $\gamma$ for the two absorbers with the lowest gallium content seems to plateau at higher excitations, which would suggest that at this level of excitation most of the charged defects have been screened and the potential fluctuations flattened. For completeness, we point out, that the same analysis of $\gamma$ was carried out in $CuInSe_2$ polycrystalline absorbers (without alloy disorder and thus without band gap fluctuations) grown by co-evaporation onto soda lime glass. We obtained similar results, supporting the origin of the potential fluctuations being electrostatic in this kind of absorbers (see section S3 of the supporting information). Reports on the more severe effect of electrostatic potential fluctuations than band gap fluctuations on device performance can be found in literature[30].

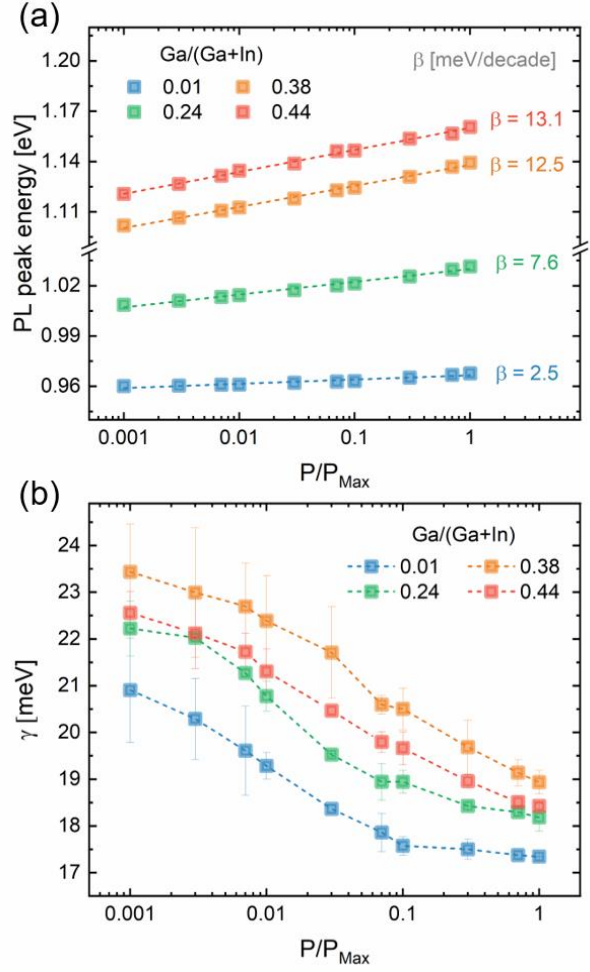

**Figure 4.** Determination of the PL peak energy position as a function of excitation. Dashed lines are linear fits, the slopes of which are used to determine β (a). Magnitude of the potential fluctuations γ determined from spectral fits to the low-energy wing of the PL at 10 K as a function of the normalized laser power. Error bars represent maximum and minimum values obtained from the fits (b).

From the experiments and analyses presented in this section, it is possible to conclude that the magnitude of the potential fluctuations in Cu(In,Ga)Se$_2$ increases with the gallium content (concluded from the composition dependence of $T_{min}$ and $\beta$), and that such fluctuations are mostly of electrostatic nature (concluded from the excitation intensity dependence of $\gamma$). Thus both: the density of tail states in these samples (values of $E_U$ in Table 1), as well as the electrostatic potential fluctuations increase with the gallium content. Additionally, in the literature, it has also been demonstrated for polycrystalline films that the magnitude of the potential fluctuations determined from low temperature photoluminescence measurements ($\beta$), correlates with the room temperature Urbach energy determined from external quantum efficiency.[50] Thus, a causal relationship can be established: The larger density of tail states in Cu(In,Ga)Se$_2$ is caused by stronger electrostatic potential fluctuations.

**2.3. Tail states and doping in single crystals**

Depending on the alkali or combination of alkalis used, the beneficial effects of postdeposition treatments on the bulk properties of polycrystalline Cu(In,Ga)Se$_2$, have been reported to yield a decreased density of tail states.[2, 17, 24] For example, when comparing polycrystalline absorbers grown with and without a Na-blocking layer, the decrease in Urbach energy measured by transient photocapacitance spectroscopy has been reported to be as large as 6 meV[51].

In order to investigate how alkalis decrease the density of tail states and whether this effect is exclusive to polycrystalline absorbers, the Urbach energy of alkali-treated Cu(In,Ga)Se$_2$ single crystals with similar copper and gallium contents (CGI ~ 1.0 and GGI ~ 0.25) is studied in the following experiments. Furthermore, besides samples grown by metalorganic vapor phase epitaxy (MOVPE), absorbers grown by molecular beam epitaxy (MBE) are also included. The MBE samples were prepared by a similar process as the films that yield the current record efficiency of 20% for single-crystal Cu(In,Ga)Se$_2$ solar cells.[26] For each set of samples, three conditions are compared; an alkali-free reference, a sample treated with only NaF and a sample treated with only KF.

First of all, the quasi-Fermi level splitting difference ($\Delta qFls$) was determined between the alkali-treated sample and an alkali-free reference, which can be seen in **Figure 5** (a). It is worth mentioning that the MBE-grown sample received NaF during the growth and not as a postdeposition treatment, however, the improvement observed in qFls is similar in both kinds of samples and comparable to the one reported for alkali-free polycrystalline absorbers treated with NaF of around 80 meV.[38] In the case of KF, both samples received it as a postdeposition treatment. In this case, a considerably lower improvement of the qFls is observed (especially in the MOVPE-grown single crystal), which goes along with our previous results on the limited effect of KF-PDTs in Cu(In,Ga)Se$_2$ single crystals with Cu-contents close to stoichiometry.[25] We attribute the larger improvement in qFls obtained with NaF to the more effective diffusion of sodium in the bulk crystal than potassium. Experimentally, it has been demonstrated by means of secondary-ion mass spectrometry that the bulk and grain boundary diffusion coefficients of sodium in Cu(In,Ga)Se$_2$ are higher than for heavier alkali elements.[52-53] In the case of diffusion in the bulk through point defects, theoretical calculations also predict a higher diffusivity of sodium, since the formation energy of alkali-related defects like Alk$_{Cu}$ and Alk$_{In}$ is lower for lighter alkali metals.[54] Nevertheless, the fact that $\Delta qFls$ is larger for both alkalis in the MBE-grown absorber may suggest an enhanced diffusion mechanism of alkalis in these kinds of samples. It has been demonstrated that the diffusion behavior of sodium in CuInSe$_2$ single crystals, for example, is affected by the density of extended defects which varies depending on the growth method.[55] Since both epitaxial Cu(In,Ga)Se$_2$ films grown on GaAs by MOVPE and MBE are known to suffer from a high density of dislocations[56-57], the enhanced diffusion of alkalis in the MBE-grown crystals could be the result of a higher density of dislocations in these absorbers.

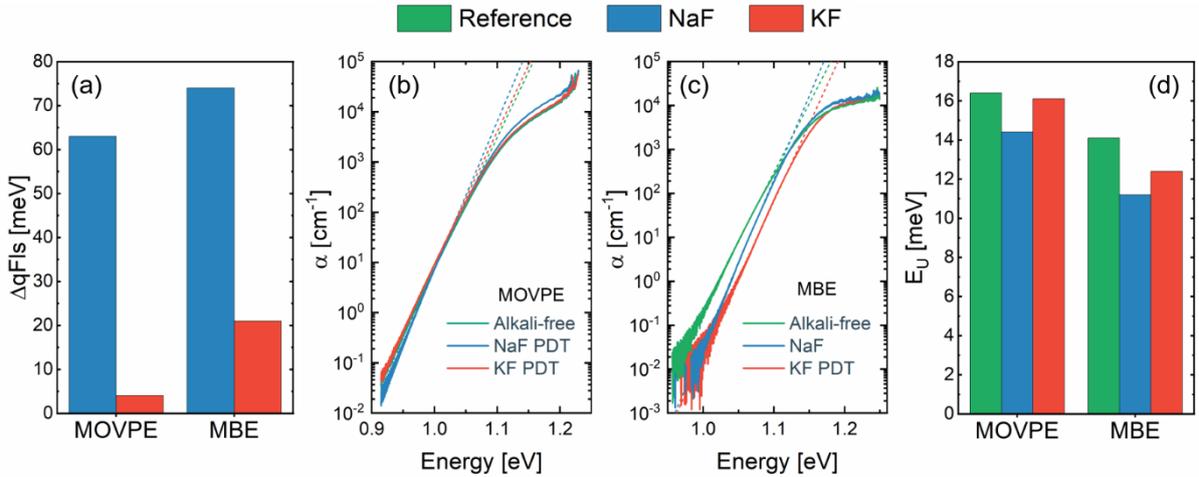

**Figure 5.** Quasi-Fermi level splitting difference ($\Delta qFls$) between alkali-treated (NaF or KF) and alkali-free Cu(In,Ga)Se$_2$ single crystals (a). Extracted absorption coefficients of all conditions for MOVPE (b) and MBE-grown samples (c) as well as the obtained Urbach energy (d).

Once the beneficial effect of alkali incorporation in the studied set of samples has been established, we proceed to study the effect of alkalis on the Urbach energy. The absorption coefficient, from which $E_U$ is extracted, of the MOVPE and MBE-grown samples for all different conditions can be seen in Figure 5 (b) and (c), respectively. From these figures, it is possible to appreciate that the steepness of the absorption onset changes with the different treatments, especially in the MBE-grown absorbers, which goes along with the more pronounced change in quasi-Fermi level splitting. As a result, variations in the Urbach energy were measured depending on the alkali metal incorporated, which are summarized in Figure 5 (d). Interestingly, the decrease in Urbach energy for the NaF-treated samples when compared to the alkali-free reference ($\Delta E_U$) of 2.8 meV (MBE) and 2.0 meV (MOVPE), is larger than the reported[2] change in state-of-the-art polycrystalline absorbers for the same comparison ($\Delta E_U$ = 0.8 meV), but similar to the reported combined effect of NaF + RbF PDT ($\Delta E_U$ = 1.9 meV)[2].

In the case of KF-treated samples, $\Delta E_U$ was found to decrease less than with NaF ($\Delta E_U$ almost negligible in the MOVPE-grown absorber), in agreement with the trend observed in the $\Delta qFls$ measured. It is thus important to highlight that the relationship between a lower Urbach energy and a lower $V_{OC}$ loss (measured as a higher quasi-Fermi level splitting), also holds true in the studied single crystals.

When taking a closer look at the relationship between the increase in quasi-Fermi level splitting and the decrease in Urbach energy, the question on the causal relationship between these two phenomena arises: is the increase in qFls a consequence of the reduced tail states or do they have a common cause? In the case of polycrystalline Cu(In,Ga)Se$_2$ absorbers, the beneficial effect of alkali postdeposition treatments on the $V_{OC}$ has been associated with a reduction of tail states in the bulk and, in particular, at the grain boundaries.[17] Evidence of reduced band-bending and accumulation of heavy alkali elements at grain boundaries has paved the way to establish the premise that both effects are correlated to the reduction in tail states measured by a decrease in the Urbach energy.[17, 23] Nonetheless, the fact that similar changes are observed in the Urbach energy of alkali-treated single crystals, where there are no grain boundaries, suggests that another mechanism related to the grain interiors also plays a role in the effect of alkalis on the Urbach energy.

From the previous analysis, it becomes evident that alkali incorporation into single crystals has the ability to reduce potential fluctuations and tail states, which can also be seen in the changes in $T_{min}$ and $\gamma$ of the MBE and MOVPE samples. On average, the magnitude of the potential fluctuations decreases by 4 meV and the redshift-blueshift behavior becomes barely noticeable upon sodium incorporation (see section S4 of the supporting information). Hence, the question on how alkali incorporation reduces tail sates can then be raised. One of the well-known effects of alkalis in Cu(In,Ga)Se$_2$, especially sodium, is to increase the net doping.[25, 58-61] Can a doping effect explain the observed changes in tail states? An immediate effect of an increased net doping level, i.e. either increased acceptor or reduced donor concentration, would be the reduction of the degree of compensation $K = N_D/N_A$ (in a p-type semiconductor), where $N_{A(D)}$ is the acceptor (donor) concentration. As we have highlighted throughout the text, tail states in Cu(In,Ga)Se$_2$ are mainly caused by electrostatic potential fluctuations, meaning that a decrease in the degree of compensation would have a direct influence on the density of charged defects and tail states. Theoretically, the average amplitude of the potential fluctuations can be estimated by equation (1)[62], which takes into account the screening of the charges and depends on the density of randomly distributed charged defects $N_C = N_A + N_D$, the carrier concentration $p$ and the material's relative permittivity $\varepsilon_r$.

$$\gamma_p = \frac{e^2}{4\pi\varepsilon_0\varepsilon_r}\frac{N_C^{2/3}}{p^{1/3}} \quad (1)$$

Decreased compensation would either decrease $N_C$ or increase $p$ and thus reduce the amplitude of the potential fluctuations. By assuming $N_C = 10^{17}\ cm^{-3}$ and $p = N_C[(1-K)/(1+K)]$ in equation (1), the magnitude of the potential fluctuations as a function of the degree of compensation was calculated and is displayed in **Figure 6**. The values of $\gamma$ obtained so far in alkali-free samples (> 21 meV), would suggest that the samples have a degree of compensation of minimum 95%, which is in excellent agreement with experimental findings determined from Hall measurements in CuGaSe$_2$.[36] The decrease of ~4 meV in $\gamma$ measured after Na incorporation in a material with $K = 0.97$, for example, would indicate that the degree of compensation has decreased by just 2-3%, demonstrating that even small changes in doping can result in diminished electrostatic potential fluctuations, and thus, in a lower density of tail states.

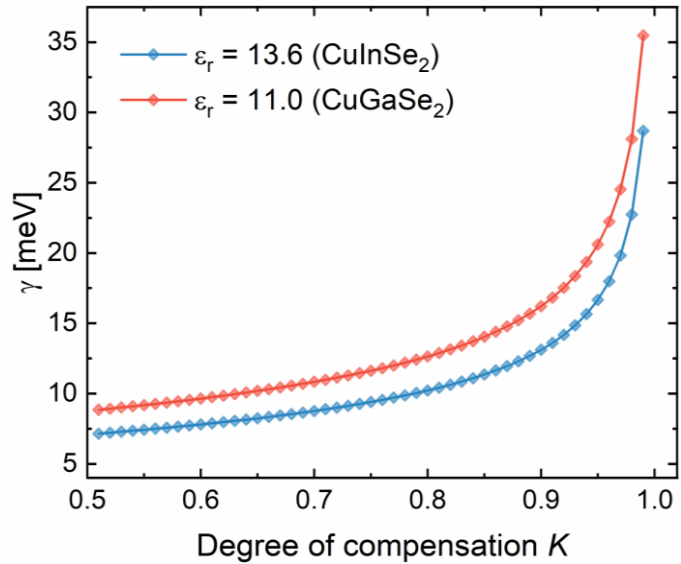

**Figure 6.** Magnitude of the potential fluctuations $\gamma$ as a function of the degree of compensation $K$ for two different dielectric constants.

The dependency of electrostatic potential fluctuations on the dielectric constant of the material, can provide a viable explanation to the observed increase in Urbach energy with band gap energy discussed in section 2.1. Based on simple arguments on bond length and electron density it can be argued that the band gap energy has an inversely proportional relation with the dielectric constant, which is found experimentally for a large class of semiconductors.[63-64] This relationship would suggest that a semiconductor with a larger band gap (and thus a smaller dielectric constant), would suffer less screening and thus from stronger electrostatic potential fluctuations. In the case of Cu(In,Ga)Se$_2$, the dielectric constant decreases from 13.6 in pure CuInSe$_2$[65] to 11[66-67] in CuGaSe$_2$, which would indicate that tail states arising from electrostatic potential fluctuations in the solid solution, would increase intrinsically with the Ga/(Ga+In) ratio.

In this section we showed, that alkali incorporation into single crystals results in a decreased density of tail sates as similarly observed in polycrystalline absorbers, suggesting that the beneficial effects of alkali PDTs on the V$_{OC}$ are not exclusively grain-boundary related. Furthermore, we showed that an increase in doping caused by the alkalis could explain the decreased density of tail states, as a decrease in the degree of compensation lowers the magnitude of electrostatic potential fluctuations. Moreover, the fact that doping decreases tail states is a direct evidence of the electrostatic nature of the potential fluctuations, since band gap fluctuations would not be affected by doping changes. In the following section, a detailed analysis on the relationship between Urbach energy and V$_{OC}$ loses, taking doping into account, is presented.

## 2.4. Tail states and V$_{OC}$ losses

Since sodium proved to be the most effective alkali in altering both quasi-Fermi level splitting and Urbach energy, a systematic study on the effect of sodium on these parameters was carried out. For the study, similar pieces of the same close-to-stoichiometric alkali-free Cu(In,Ga)Se$_2$ single crystal with a Ga/(Ga+In)~0.4 were cleaved and treated with a NaF-PDT with varying annealing time (to have the same crystal with varying sodium content). To prevent variations in the amount of sodium fluoride before the annealing step, all samples were loaded into the evaporation chamber and received NaF at the same time. **Figure 7** (a) shows the quasi-Fermi level splitting at 1 sun (details can be found in section S5 of the supporting information) and the extracted Urbach energies of the studied samples including an alkali-free reference (denoted as zero minutes annealing time). This figure demonstrates the direct correlation between the increase of sodium in the film and the decrease in Urbach energy and increase of quasi-Fermi level splitting. Since changes in the splitting of the Fermi level can arise from both electron ($E_{F,e}$) and hole ($E_{F,h}$) quasi-Fermi levels due to changes in the recombination activity and doping concentration[68], respectively, we will discuss the role of $E_U$ on the observed changes in the following.

Wolter et al.[2] addressed the $V_{OC}$ losses associated with recombination through tail states with respect to the Shockley-Queisser limit. Since tail states act as non-radiative recombination centers but also allow absorption below the band gap[9], the model takes into account the effect in both radiative and non-radiative voltage loss. According to the model, an additional $V_{OC}$ loss of around 10 mV would be expected for every meV increase in $E_U$ as a consequence of the increased radiative and non-radiative recombination from tail states (each effect contributing roughly 5 mV). However, the model can only explain half of the experimentally observed $V_{OC}$ loss, suggesting that another mechanism related to tail sates needs to be further taken into account.[2] In the following, we analyze the $V_{OC}$ loss by studying the changes in quasi-Fermi level splitting, since $V_{OC} = qFls/q$ in the absence of gradients in the quasi Fermi levels at open circuit.[69] Moreover, it has been demonstrated that the $V_{OC}$ almost equals the qFls in high-efficiency Cu(In,Ga)Se$_2$ devices[38].

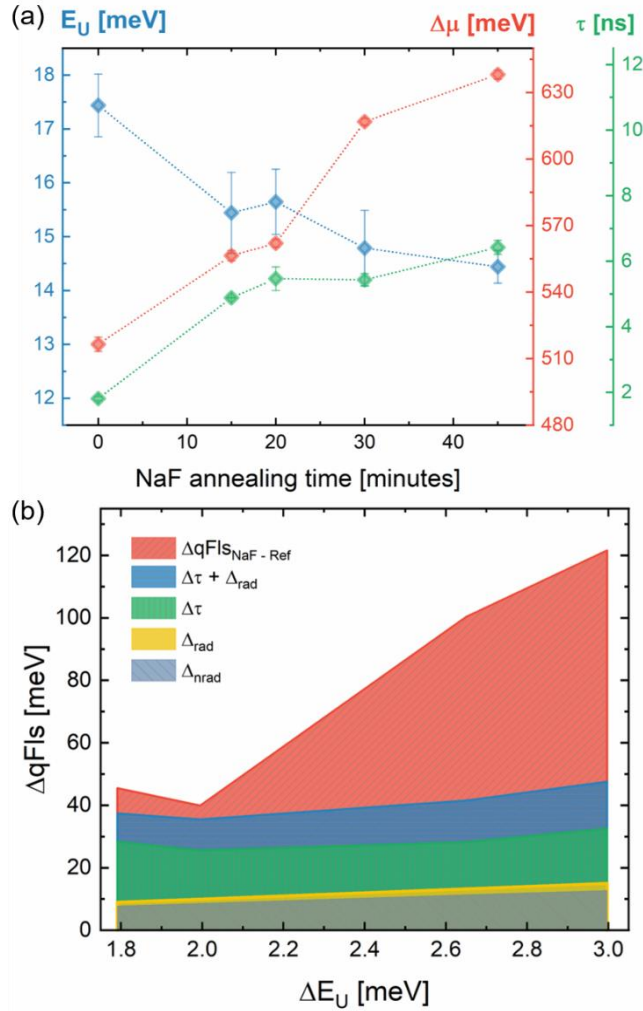

**Figure 7.** Urbach energy, quasi-Fermi level splitting and minority carrier lifetime of Cu(In,Ga)Se$_2$ single crystals treated with NaF and varying annealing time. A sodium-free reference is also included and denoted as zero minutes annealing time (a). Area plot (areas start from zero) of the contributions to the quasi-Fermi level splitting difference between the Na-containing and Na-free reference sample as a function of the decrease in Urbach energy (note that $\Delta E_u$ does not start from zero, since we study the changed to the Na-free reference) (b).

Focusing on the non-radiative loss due to Shockley-Read-Hall recombination, the model proposes that the change in V$_{OC}$ due to a change in Urbach energy can be quantified as:

$$q\Delta\Delta Voc^{nrad} \approx \Delta\Delta qFls^{nrad} \approx k_B T \ln\left(\frac{E_R^B}{E_R^A}\right) \qquad (2)$$

Where $k_B$ is the Boltzmann constant, $T$ the temperature in Kelvin, $q$ the elementary charge and $E_R$ the decay energy of the Shockley-Read-Hall recombination rate due to recombination through tail states defined as:

$$\frac{1}{E_R} = \frac{1}{E_U} - \frac{1}{k_B T} \qquad (3)$$

From the determined Urbach energy values, $E_R$ was calculated for all conditions with the use of equation 3. Then, the change in quasi-Fermi level splitting $\Delta\Delta qFls$ due to the non-radiative loss caused by recombination through tail states was calculated from equation (2). For example, by taking the Na-free sample as reference ($E_R = 54\ meV$) and the one with the highest amount of sodium ($E_R = 33\ meV$), an increase in quasi-Fermi level splitting of roughly 13 meV due to a decrease of 3 meV in Urbach energy would be expected. Since this change is caused by a decrease in non-radiative recombination, and thus related to the electron quasi-Fermi level $E_{F,e}$, a corresponding change in minority carrier lifetime should also be observed. In order to compare the theoretically calculated non-radiative loss with the one determined from changes in lifetime, we performed time-resolved photoluminescence measurements (details can be found in section S6 of the supporting information). As can be seen from Figure 7 (a), the minority carrier lifetime follows the increasing trend in quasi-Fermi level splitting, confirming that the observed improvement is caused partially due to the effectiveness of sodium in reducing non-radiative recombination centers such as tail states. From the measured minority carrier lifetimes, we proceeded to quantify the change in qFls due to $E_{F,e}$ as[69]:

$$\Delta qFls_{(E_{F,e})} = k_B T \ln\left(\frac{\tau_{NaF}}{\tau_{Ref}}\right) \quad (4)$$

Where $\tau_{NaF,Ref}$ is the minority carrier lifetime of the Na-containing and alkali-free samples, respectively. By taking the same samples as in the previous example, a change in quasi-Fermi level splitting of 32 meV would be expected from the increase in lifetime from 1.8 ns (alkali-free) to 6.4 ns (most Na-containing). Interestingly, the change in lifetime is larger than the one expected from just the reduction in non-radiative recombination due to reduced tail states. This would indicate that the effect of sodium in reducing non-radiative recombination centers in these samples extends beyond tail states, as will be further demonstrated below.

As already mentioned, the model also takes into account the radiative $V_{OC}$ loss caused by tail states, which is estimated to be around 5 mV per each meV increase in Urbach energy.[2] To have a better visualization of the impact of the loss mechanisms on the change in quasi-Fermi level splitting as a function of the decrease in Urbach energy, the area plot in Figure 7 (b) was constructed. The larger area in red denotes the measured change in quasi-Fermi level splitting between the NaF-treated samples and the alkali-free reference (all areas start from zero). The smallest gray and yellow areas account for the reduction in non-radiative ($\Delta_{nrad}$) and radiative loss ($\Delta_{rad}$) due to the reduction in Urbach energy, the green area represents the contribution $\Delta qFls$ ($\Delta_\tau$) due to the measured increase in lifetime and the blue area is the sum of the previous two ($\Delta_\tau + \Delta_{rad}$). Since the sum of $\Delta_\tau$ and $\Delta_{rad}$ cannot account for the whole improvement in quasi-Fermi level splitting, we attribute the remaining proportion to the contribution from $E_{F,h}$ caused by the well-known effect of sodium discussed in the previous section: doping. In fact, the double effect of sodium as dopant and effective passivator of non-radiative recombination centers has also been observed in polycrystalline Cu(In,Ga)Se$_2$ absorbers, where a continuous increase in hole carrier concentration accompanied by changes in the defect spectra detected by admittance spectroscopy and deep level transient spectroscopy were measured as a function of the sodium content[70].

From the $\Delta qFls$ difference between the red and blue area in Figure 7 (b), it is possible to see that the contribution to the quasi-Fermi level splitting improvement from doping $\Delta qFls_{(E_{F,h})}$, would range between 8 to 75 meV depending on the NaF annealing time. With the help of these values and equation (5), the hole carrier concentration can be estimated to increase by a factor of 1.4 to 18.5 as the annealing time increases. It is worth mentioning that reports of increased doping after alkali postdeposition treatments of the same magnitude as the estimated one can be found in literature.[58, 70-71] Furthermore, Figure 7(b) also shows that the initial decrease in Urbach energy ($\Delta E_U = 1.8 - 2\ meV$ corresponding to 15 and 20 minutes annealing time) is mostly associated to the reduction of non-radiative recombination centers with just a small contribution from doping. However, as discussed in the previous section, this seemingly small change in doping can have a strong effect on electrostatic potential fluctuations, and hence on tail states, depending on the degree of compensation. Thus, from this analysis it is possible to conclude that the $V_{OC}$ improvement obtained from doping, can account for the missing contribution in the model of radiative and non-radiative recombination through tail states to fully describe the entirety of the voltage losses in Cu(In,Ga)Se$_2$.

$$\Delta qFls_{(E_{F,h})} = k_B T\ ln\left(\frac{p_{NaF}}{p_{Ref}}\right) \qquad (5)$$

To better visualize the combined effect of doping, the radiative and the non-radiative voltage loss caused by tail states, a comparison between high-efficiency polycrystalline and single crystalline solar cells was carried out. In the case of single crystals, devices with alkali-free, NaF, KF and NaF+KF absorbers were fabricated by MBE. The Urbach energy was extracted from photoluminescence measurements under one sun illumination. The values of $E_U$ and the device parameters can be seen in Table 2. In the case of polycrystalline absorbers, data were taken from reference [2] and include similar devices with different alkali PDTs such as NaF and NaF+RbF. In order to set aside differences in the band gap, the $V_{OC}$ loss with respect to the Shockley-Queisser $V_{OC}$ was calculated. In the case of the single crystals, the band gap determined from the maximum of the PL spectra was 1.15 eV, which has a corresponding Shockley-Queisser $V_{OC}$ of 906 mV[45] (with back reflector). With all this information, the voltage loss was calculated and is displayed in **Figure 8** as a function of the Urbach energy.

**Table 2.** Urbach energy and device parameters of single crystalline solar cells with different alkali treatments (area around 0.2 cm$^2$).

| Alkali treatment | $E_U$ [meV] | $V_{OC}$ [mV] | $J_{SC}$ [mA/cm$^2$] | FF | $\eta$ [%] |
|---|---|---|---|---|---|
| None | 13.0 ± 0.2 | 668 | 33.2 | 0.75 | 16.6 |
| NaF | 12.2 ± 0.3 | 727 | 34.2 | 0.781 | 19.4 |
| KF | 12.9 ± 0.1 | 686 | 34.4 | 0.768 | 18.1 |
| NaF+KF | 11.8 ± 0.4 | 754 | 34.5 | 0.815 | 21.2 |

The blue area in Figure 8 corresponds to the expected radiative and non-radiative voltage loss due to $E_U$ alone, while the grey area accounts for the $V_{OC}$ improvement obtained as a consequence of an increase in doping. For example, a decrease in Urbach energy from 16 to 11 meV, would convey a reduced voltage loss of roughly 100 mV. If 50 of those 100 mV were caused by an increased doping (which would imply that the doping increases by a factor of ~7), the remaining 50 mV could be explained by the reduced radiative and non-radiative voltage loss caused by the diminished density of tail states. Interestingly, both polycrystalline and single crystalline devices that contain sodium (either alone or in combination with a heavier alkali), lie close to the region where the voltage loss can be explained entirely by tail states. This further speaks in favor of tail states determined mainly by grain interior properties rather than grain boundaries. Furthermore, the data points that deviate the most correspond to devices that do not contain sodium, which would indicate that the double effect of sodium as dopant and passivator of non-radiative recombination centers would also be necessary in order to reduce the voltage loss. If upon sodium incorporation the $V_{OC}$ loss would still be far from the tails-limited region, this would suggest that another source of non-radiative recombination is the origin of the loss, such as a deficient passivation of the interfaces or the presence of deep defects. The additional improvement of KF treatment after NaF treatment has been attributed to surface passivation[26]. Our data confirms this explanation, as the NaF+KF epi sample moves closer to the line where the $V_{OC}$ loss is entirely due to tail states. Furthermore, the additional KF further reduces the Urbach energy by increasing the doping further and reducing the electrostatic potential fluctuations, but to a lesser degree than NaF alone.

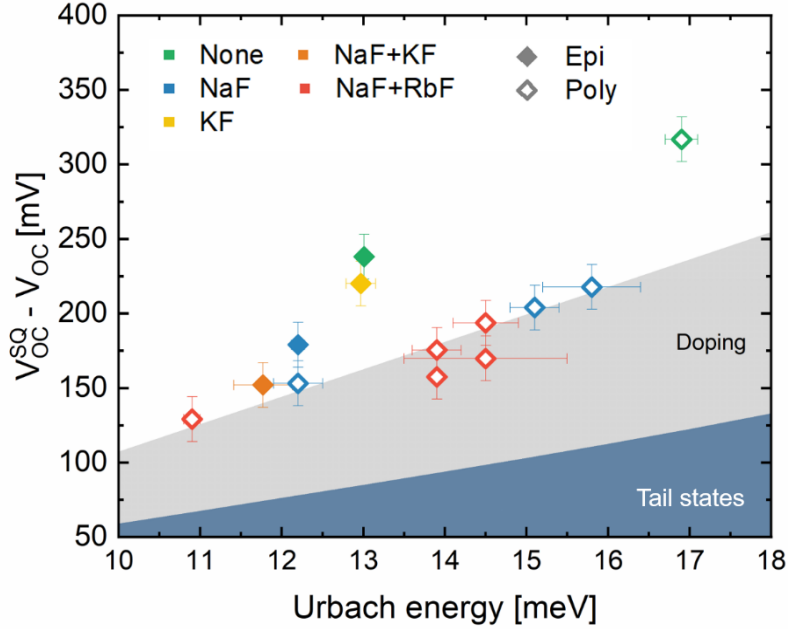

**Figure 8.** Voltage loss with respect to the Shockley-Queisser $V_{OC}$ as a function of the Urbach energy of polycrystalline and single crystalline high-efficiency Cu(In,Ga)Se$_2$ solar cells. The devices were fabricated from absorbers that received none or different alkali postdeposition treatments. Data points corresponding to polycrystalline absorbers were taken from reference [2].

Interestingly, the device with the lowest $V_{OC}$ loss ($E_U \approx 11\ meV$) has a considerably lower band gap energy of 1.08 eV[2] compared to the conventional 1.15 eV of 3-stage co-evaporated Cu(In,Ga)Se$_2$. Furthermore, the current record efficiency of 23.35% is also based on an absorber with a band gap of 1.08 eV.[20] If the lowest density of tail states achievable in an absorber is capped by the band gap energy as discussed in section 2.1, a decrease in band gap from 1.15 to 1.08 eV would allow a decrease in Urbach energy of roughly 1.4 meV (see Figure 2 (b)), which would translate into a reduced voltage loss of approximately 28 meV. Thus, it is reasonable to assume that the reduced band gap energy in these high-efficiency devices is at least partially responsible for the reduced Urbach energy and thus the reduced voltage loss, and a proven strategy to follow up in order to continue boosting the efficiency of Cu(In,Ga)Se$_2$.

All in all, the results presented in this section shed light on the beneficial effect of alkalis, and especially sodium, in reducing the $V_{OC}$ deficit of Cu(In,Ga)Se$_2$ solar cells. The role of sodium could explain why the reported improvement in $V_{OC}$ obtained after a postdeposition treatment, for example, varies from laboratory to laboratory. If the properties of the substrate, back contact and growth-process itself allowed an adequate sodium diffusion, then the improvement after a PDT would be rather small. On the contrary, if the initial sodium concentration was not optimal, then the beneficial effect of the PDT would be greater. Evidence of this can be found in literature, where the improvement after a KF PDT in polycrystalline absorbers grown on soda lime glass has been reported to span from 20 mV[72] to 65 mV[73].

The effects of alkali postdeposition treatments on the $V_{OC}$ improvement of Cu(In,Ga)Se$_2$ solar cells, has long been associated with grain boundary passivation measured by changes in band bending at grain boundaries.[17, 23] However, our results suggest that rather than grain boundaries, it is the increased sodium concentration at the grain interiors that is the driver of the $V_{OC}$ improvement. Nevertheless, in order to increase the sodium content in the grains of polycrystalline absorbers, a heavier alkali species is necessary to "push" sodium out of the grain boundaries into the grain interiors[23], explaining why the effect of postdeposition treatments on the $V_{OC}$ is similar regardless of the heavy alkali species used. Furthermore, the validity of the observed changes in band bending at grain boundaries has recently been questioned by demonstrating the strong effect of surface roughness on the accuracy of Kelvin probe force microscopy measurements.[74] The conclusion that the improved $V_{OC}$ is due higher Na in the grains rather than grain boundary passivation is also in agreement with the observation that alkali treatment does not reduce the recombination velocity at grain boundaries[75].

## 3. Conclusion

In this contribution, we present a photoluminescence study on tail states in Cu(In,Ga)Se$_2$ single crystals, which we quantify by their Urbach energy, and how it is affected by compositional changes and the incorporation of alkali metals. Our results suggest that not only the copper but also the gallium content has an influence on the Urbach energy, which increases for higher gallium contents. A temperature and intensity-dependent photoluminescence study of samples with different Ga/(Ga+In), reveals that samples with higher gallium contents suffer from more harmful potential fluctuations. Contrary to what is expected from the literature, the increased Ga content does not seem to result in fluctuations arising from compositional inhomogeneities (band gap fluctuations), but to yield electrostatic potential fluctuations.

Regarding the alkali incorporation, our results unveiled that the Urbach energy of MOVPE and MBE-grown single crystals is affected in a similar way as in the case of polycrystalline absorbers. In the case of Na-containing crystals, the Urbach energy was found to decrease when compared to an alkali-free reference: 2.8 meV for the MBE and 2.0 meV in the MOVPE-grown sample. In the case of KF, only the MBE-grown sample showed a significant decrease in $E_U$ of around 1.7 meV. In all samples, the improvement in quasi-Fermi level splitting with the alkali treatment was found to be correlated with the change in the Urbach energy, confirming that the empirical relationship of a lower Urbach energy with a lower $V_{OC}$ deficit also applies in the studied single crystals.

Furthermore, we carried out an analysis of the voltage losses in single crystals, where we show that upon alkali incorporation the doping concentration increases, leading to an increase in quasi-Fermi level splitting and a reduction in the density of tail states. We explain that as a consequence of the increased doping concentration, the degree of compensation decreases, resulting in a lower density of charged defects that are the cause of electrostatic potential fluctuations, which are the origin of tail states in $Cu(In,Ga)Se_2$. Moreover, we argue that the $V_{OC}$ improvement that could not be explained through radiative and non-radiative recombination via tails states can be explained by the effect of doping. Higher doping leads to lower Urbach energies via reduced electrostatic potential fluctuations and at the same time increases $V_{OC}$ or qFls.

Finally, by comparing the $V_{OC}$ loss with respect to the Shockley-Queisser $V_{OC}$ in polycrystalline and single crystalline solar cells treated with different alkalis, we concluded that tail states are determined by the grain interior properties rather than by grain boundaries, since the $V_{OC}$ loss of sodium-containing devices was found similar in both kinds of absorbers. This contribution demonstrates that an increase in sodium at the grain interiors has a strong effect on the optoelectronic properties of $Cu(In,Ga)Se_2$, and suggests that such increase is the main driver of the $V_{OC}$ improvement obtained after a postdeposition treatment with KF, RbF or CsF in polycrystalline absorbers rather than grain boundary passivation.

## 4. Experimental Section

*Metalorganic vapor phase epitaxy (MOVPE)*: $Cu(In_{1-x},Ga_x)Se_2$ single crystals were grown on 500μm thick (100)-oriented semi-insulating GaAs wafers. The reactor temperature and pressure were set at 520°C and 90mbar for all the processes. The metalorganic precursors used were cyclopentadienyl-coppertriethyl phosphine (CpCuTEP), trimethylindium (TMIn), triethylgallium (TEGa) and diisopropylselenide (DiPSe). In order to achieve the desired composition, the CpCuTEP partial pressure was kept constant and the partial pressures of TMIn and TEGa adjusted for each process. Selenium is always provided in excess.

*Postdeposition treatments in MOVPE samples*: After growth, each sample was cleaved in a $N_2$-filled glovebox and transferred into an MBE chamber where potassium fluoride was evaporated at a rate of roughly 1nm/min during 6 minutes. The KF deposition was done under a Se pressure of ~2x10$^{-6}$ Torr at a substrate temperature of 350°C. After the treatment, the samples were rinsed with DI water and etched in a 5% KCN solution for 30 seconds right before the CdS buffer layer deposition. For the NaF-PDT, each sample was cleaved in a N$_2$-filled glovebox and transferred into an electron beam evaporator where 10nm of NaF were deposited. After this, the samples went back into the glovebox to receive a first annealing step at 350°C before being transferred into an MBE chamber where they were selenized under the same conditions as for the KF-PDT.

*Photoluminescence*: Measurements were performed using a 660nm diode laser with a spot diameter of ~2.6mm as excitation source. Two parabolic mirrors were used to collect and redirect the emitted photoluminescence into a spectrometer where it was detected by an InGaAs array. All measurements for qFls and Urbach energy determination were carried out at room temperature and spectrally corrected using a calibrated halogen lamp. For the analysis performed at low temperatures, a liquid helium flow cryostat was used. For the determination of the quasi-Fermi level splitting, the value was extracted from a linear fit of the high-energy wing of the emitted PL (described by Planck's generalized law). Details on the extraction of the Urbach energy can be found in the supporting information. Time resolved photoluminescence (TRPL) measurements were taken with a LifeSpec II Spectrometer equipped with a 640nm pulsed diode laser.

*Chemical characterization:* Elemental composition was determined by energy dispersive X-ray spectroscopy (EDX) taken at an acceleration voltage of 10kV.

**Acknowledgements**

This research was funded in whole, or in part, by the Luxembourg National Research Fund (FNR), grant reference C17/MS/11696002 GRISC and PRIDE17/12246511/PACE. A. Prot thanks AVANCIS GmbH in the framework of the project POLCA. M.H. Wolter thanks the European Union's Horizon 2020 Research and Innovation Program under Grant Agreement No. 641004 (Sharc25). For the purpose of open access, the author has applied a Creative Commons Attribution 4.0 International (CC BY 4.0) license to any Author Accepted Manuscript version arising from this submission.

**Supporting Information**

**On the origin of tail states and V$_{OC}$ losses in Cu(In,Ga)Se$_2$**

*Omar Ramírez[*], Jiro Nishinaga, Felix Dingwell, Taowen Wang, Aubin Prot, Max Hilaire Wolter, Vibha Ranjan and Susanne Siebentritt*

O. Ramírez, F. Dingwell, T. Wang, A. Prot, M. H. Wolter, V. Ranjan and Prof. Dr. S. Siebentritt
Department of Physics and Materials Science, University of Luxembourg, 41 rue du Brill, L-4422, Belvaux, Luxembourg
E-mail: omar.ramirez@uni.lu

Dr. J. Nishinaga
Renewable Energy Research Center, National Institute of Advanced Industrial Science and Technology (AIST), Koriyama, Fukushima, 903-0298, Japan.

**Table S1**. Copper and gallium content, optical bandgap determined from photoluminescence and extracted Urbach energy of all the single crystals used in the present study.

| Sample No. | Name | Cu/(Ga+In) | Ga/(Ga+In) | $Eg_{PL}$ [eV] | $E_U$ [meV] |
|---|---|---|---|---|---|
| 1 | 268 | 0.95 | 0.42 | 1.26 | 18.3 |
| 2 | 270-BR | 1.04 | 0.36 | 1.23 | 15.6 |
| 3 | 271 | 1.056 | 0.02 | 1.04 | 11.6 |
| 4 | 271-2H | 1.04 | 0.21 | 1.06 | 12.2 |
| 5 | 271-3H | 1.02 | 0.25 | 1.07 | 12.7 |
| 6 | 272 | 1.02 | 0.42 | 1.25 | 16.2 |
| 7 | 273 | 1.04 | 0.43 | 1.25 | 17.5 |
| 8 | 274 | 0.97 | 0.44 | 1.26 | 17.5 |
| 9 | 275 | 1.02 | 0.38 | 1.24 | 16.8 |
| 10 | 276 | 0.95 | 0.3 | 1.15 | 16.4 |
| 11 | 280 | 1.07 | 0.03 | 1.03 | 12.1 |
| 12 | 282 | 0.98 | 0.44 | 1.26 | 17.2 |
| 13 | 285 | 1.06 | 0.01 | 1.03 | 12.6 |
| 14 | 288 | 0.99 | 0.24 | 1.12 | 15.0 |
| 15 | AIST-A | 1.0 | 0.25 | 1.14 | 14.1 |
| 16 | 438 | 0.97 | 0.0 | 1.02 | 14.8 |

## S1. Urbach energy determination from PL

In order to determine the Urbach energy, the absorption coefficient is needed, which is obtained from Beer – Lambert's law as:

$$\alpha(E) = -\frac{\ln(1 - a(E))}{d}$$

Where $d$ is the sample's thickness and $a(E)$ the absorptivity obtained from applying generalized Plank's law in Boltzmann approximation to the room temperature PL spectrum. Note that the quasi-Fermi level splitting and temperature are required in order to obtain $a(E)$. After that, the absorption coefficient is fitted with:

$$\alpha(E) = \alpha_0 \exp\left(-\frac{E_0 - E}{E_U}\right)$$

The decay energy $E_U$ is the so-called Urbach energy, which describes the exponential decay nature of the tail states into the bandgap. Fittings are performed below the bandgap and above the region where the PL spectrum exhibits considerable noise and in different energy ranges in order to reduce the error. Figure S1 exemplifies the fitting to the absorption coefficient in order to obtain the Urbach energy.

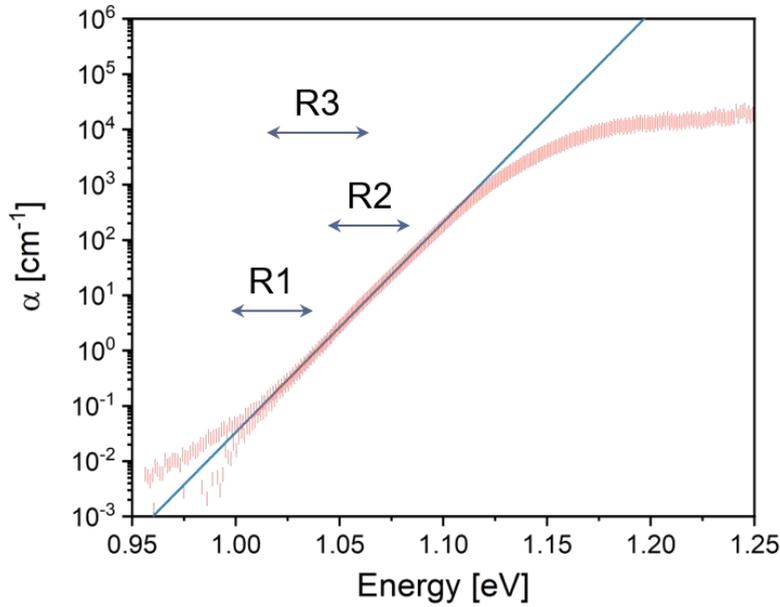

**Figure S1.** Example of a fitting routine for the determination of the Urbach energy. Different arrows denote different fitting ranges. The blue line is the resulting fit from which the Urbach energy is extracted.

## S2. Contour maps of the temperature-dependent photoluminescence

In order to create the contour maps of Figure 3, a temperature-dependent analysis of each sample was carried out in a liquid helium flow cryostat. The samples were excited with 3mW of a 660 nm diode laser and the obtained spectra were spectrally corrected using a calibrated halogen lamp. Figure S2 shows the temperature-dependent photoluminescence of the $CuIn_{0.56}Ga_{0.44}Se_2$ sample discussed in the main text. After that, all the spectra were normalized and the PL intensity set as the Z-axis.

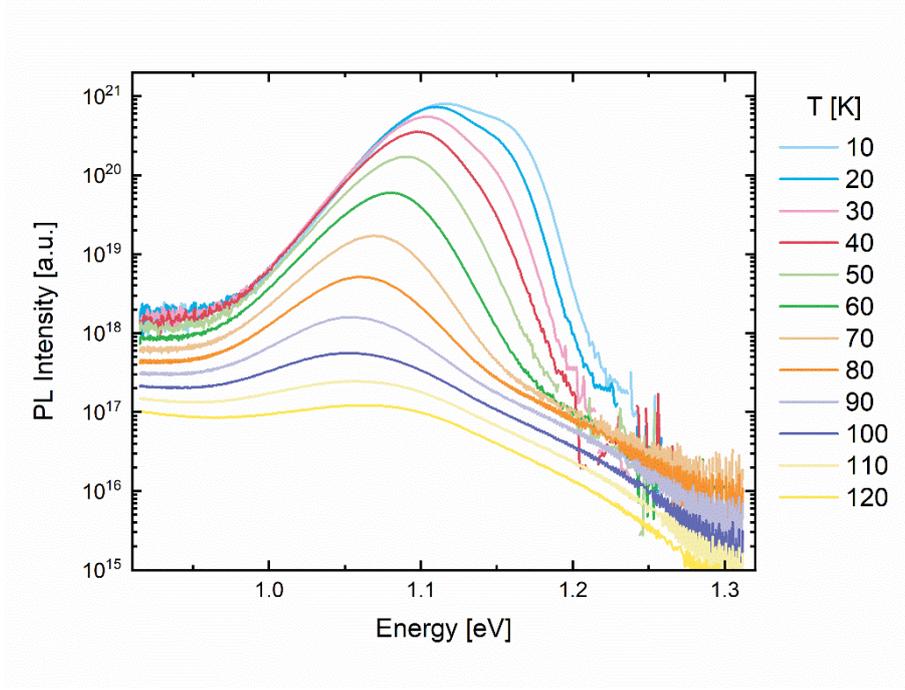

**Figure S2.** Temperature-dependent photoluminescence of an alkali-free CuIn$_{0.56}$Ga$_{0.44}$Se$_2$ single crystal.

## S3. Determination of β and γ from intensity-dependent photoluminescence

For the determination of β, intensity-dependent photoluminescence measurements at 10K were performed. By using absorptive neutral density filters, the optical density is varied allowing the attenuation of the incident laser beam. The measurements were carried out over three orders of magnitude. After that, each PL spectrum was spectrally corrected, normalized and fitted with an asymmetric double sigmoidal function to determine the energy of the peak position. The exact function has no physical meaning, it just used to obtain the energy of the maximum. It is worth mentioning that the increasing trend of β with the gallium content discussed in the manuscript is independent of the fitting method used. Taking simply the energy of the highest intensity of the PL emission, for example, will show the same trend.

For the determination of γ, the same PL spectra were used. The low energy wing of the PL was fitted with the equation:

$$I(E) = A_0 \exp\left(\frac{E}{\gamma}\right)$$

The fitting range was restricted to at least 50 meV away from the PL maximum and at least two fitting routines were carried out in different energy ranges. The values of γ presented in the main text are the average of the different values of γ obtained for the different fitting ranges. As an example, figure S3 shows the intensity-dependent photoluminescence analysis with the corresponding fits to determine β (a) and γ (b) from the sample with a Ga/(Ga+In) of 0.24 discussed in the text.

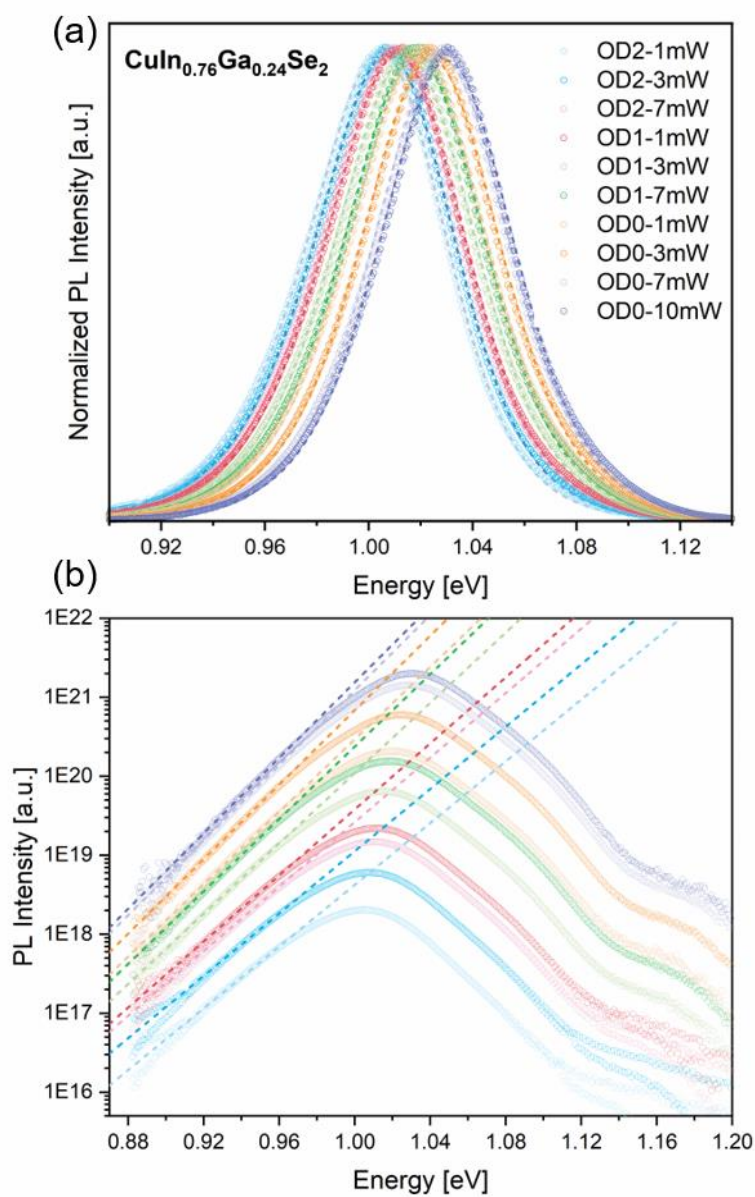

**Figure S3.1.** Excitation-dependent photoluminescence of an alkali-free $CuIn_{0.76}Ga_{0.24}Se_2$ single crystal. Dashed lines are best fits to the data for the determination of the beta parameter (a) and gamma (b). Same color code applied to both figures.

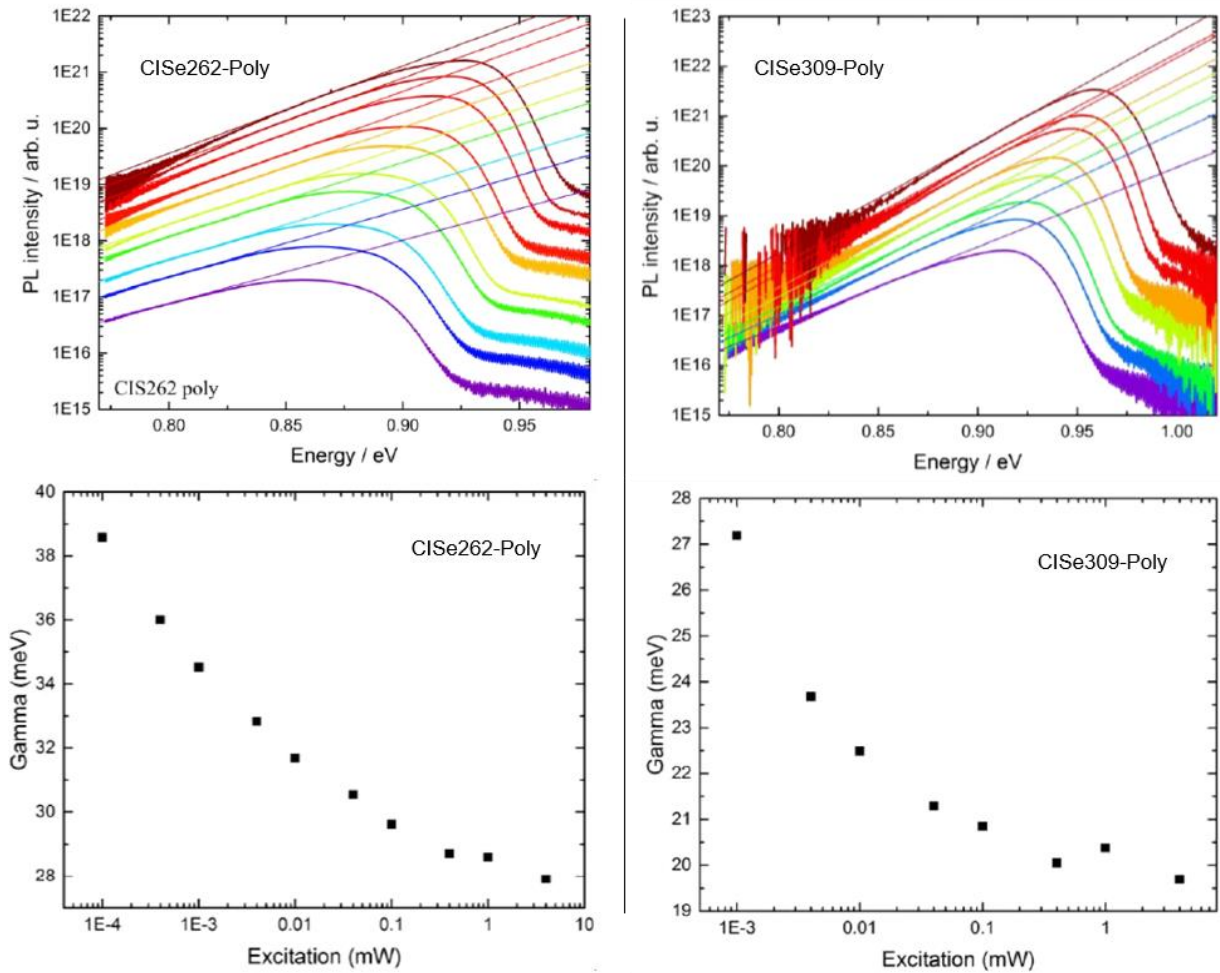

**Figure S3.2.** Excitation-dependent photoluminescence at 10K of two polycrystalline CuInSe$_2$ absorbers (up) and the magnitude of the potential fluctuations γ determined from the spectral fits as a function of the laser power (down).

**S4. Effect of sodium on T$_{min}$ and $\gamma$**

In the main manuscript, we concluded that tail states in Cu(In,Ga)Se$_2$ are caused by electrostatic potential fluctuations and that the addition of alkali metals, especially sodium, have an impact on such fluctuations which is measured as a decrease in the Urbach energy after the alkali postdeposition treatments. Here, the changes in potential fluctuations upon alkali incorporation are discussed.

Similarly to section 2.2, the magnitude of the potential fluctuations was determined from intensity and temperature-dependent photoluminescence measurements. As can be seen from Figure S4, the incorporation of sodium after the PDT resulted in a decrease in T$_{min}$ of the MOVPE-grown absorber discussed in section 2.3. This result, along with the measured decrease in Urbach energy of 2 meV, provides a strong evidence of the decrease in potential fluctuations caused by the alkalis. In this case, KF seems not to have an impact on T$_{min}$, which goes along with the almost negligible change in Urbach energy of this treatment.

Furthermore, the magnitude of the potential fluctuations ($\gamma$ parameter) and its dependency on excitation were also studied (Figure S5). In agreement with the temperature-dependent analysis, the magnitude of the potential fluctuations was found to decrease when comparing Na-containing samples to the alkali-free reference regardless of the growth method. On the other hand, in the case of the KF-treated absorbers, the $\gamma$ parameter of the MOVPE-grown sample is not greatly affected by the treatment, as similarly observed from T$_{min}$ in Figure S4, while in the case of the MBE samples treated with KF, the decrease in potential fluctuations is less but comparable to the NaF case. In both cases, the magnitude of gamma was found to decrease with excitation, supporting the electrostatic nature of the potential fluctuations.

In summary, both temperature and intensity-dependent photoluminescence measurements confirm that the reduction in Urbach energy obtained after alkali metal postdeposition treatments is caused, at least partially, by a decrease in electrostatic potential fluctuations.

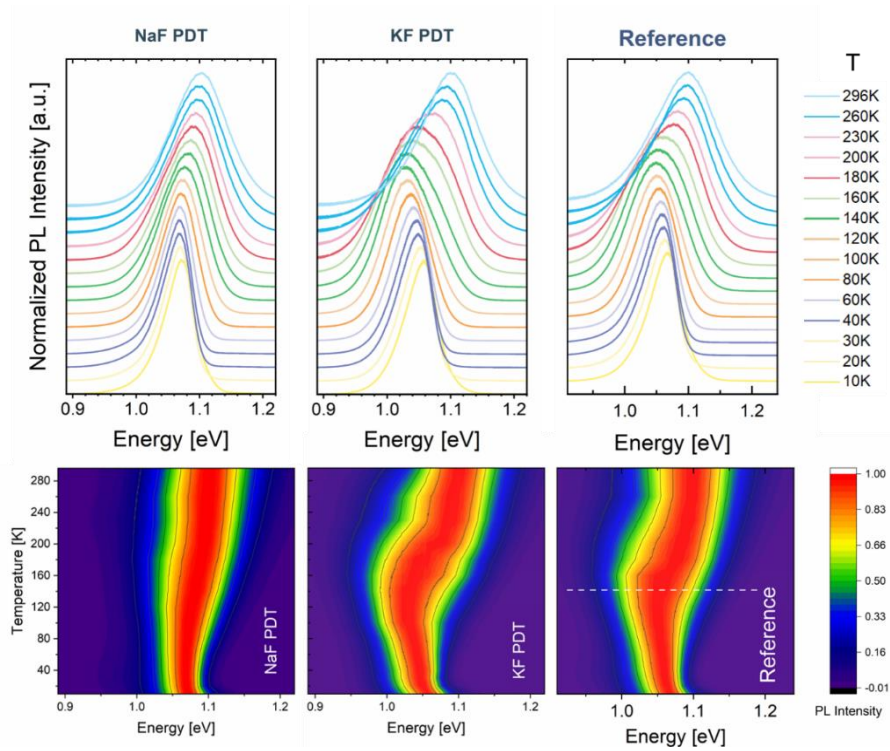

**Figure S4**. Temperature-dependent photoluminescence analysis of the MOVPE-grown single crystal with Cu content close to stoichiometry and GGI~0.24 discussed in section 2.3. From the figure is possible to appreciate the decrease in the redshift-blueshift for the NaF case (left).

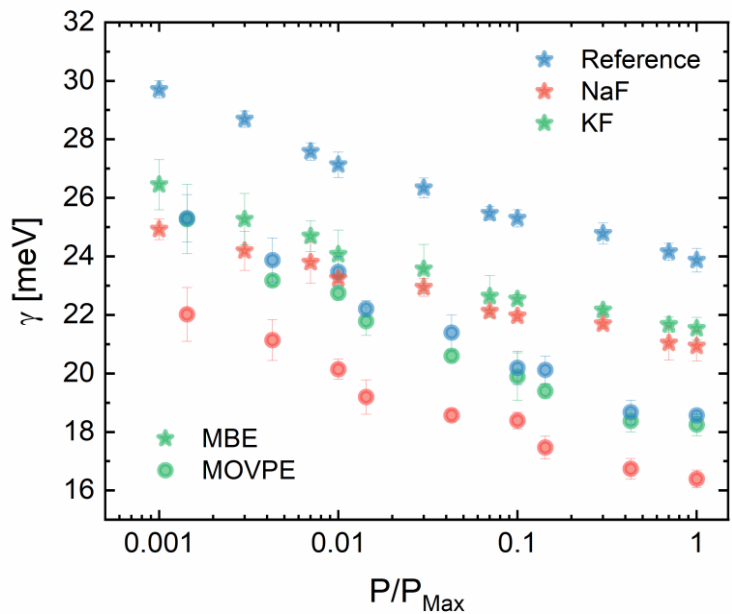

**Figure S5.** Intensity-dependence of the γ parameter of MOVPE (circles) and MBE-grown samples (stars) treated with different alkali metals.

## S5. Quasi-Fermi level splitting

In section 2.4 of the main manuscript, the quasi-Fermi level splitting values of a Cu(In,Ga)Se$_2$ single crystal with a Ga/(Ga+In)~0.4 and different sodium contents are reported. Since the alkali-free reference and low-sodium samples have considerably low PL yield, the determination of the quasi-Fermi level splitting at one sun excitation from the fit to the high-energy wing of the PL spectrum using generalized Planck's law in Boltzmann approximation is not possible. Instead, the quasi-Fermi level splitting of each sample was determined from the extrapolation of an intensity-dependent analysis carried out in a calibrated setup for photon counting.

Figure S6 (a) shows the integrated PL yield as a function of the excitation flux density. Linear fits were used to obtain the integrated PL flux equivalent to one sun illumination that, for a bandgap of 1.23 eV (from PL maximum), is 2.4x10$^{17}$ photons cm$^{-2}$s$^{-1}$. From these values, we proceeded to calculate the quasi-Fermi level splitting as $\Delta\mu = qV_{OC}^{SQ} + k_BT \ln(ERE)$. Furthermore, in order to corroborate the obtained values, the high-energy wing of the PL spectra was fitted as previously described and the quasi-Femi level splitting at one sun obtained from an extrapolation as illustrated in Figure S6 (b). The obtained quasi-Fermi level splitting from both methods were found to be in good agreement and are summarized in Table S2. The values given in the main manuscript correspond to their average.

**Table S2**. Quasi-Fermi level splitting of sample CIGSe273 treated with different NaF conditions calculated from ERE and fitting to the high-energy wing of the PL.

| Sample | $\Delta\mu_{ERE}$ [meV] | $\Delta\mu_{Fit}$ [meV] | $\Delta\mu_{avg}$ [meV] |
|---|---|---|---|
| Na-free | 519.7 | 513.3 | 516.5 |
| NaF – 15min | 558.7 | 554.1 | 556.4 |
| NaF – 20min | 563.7 | 560.2 | 562.0 |
| NaF – 30min | 618.4 | 615.3 | 616.9 |
| NaF – 45 min | 640.1 | 636.1 | 638.1 |

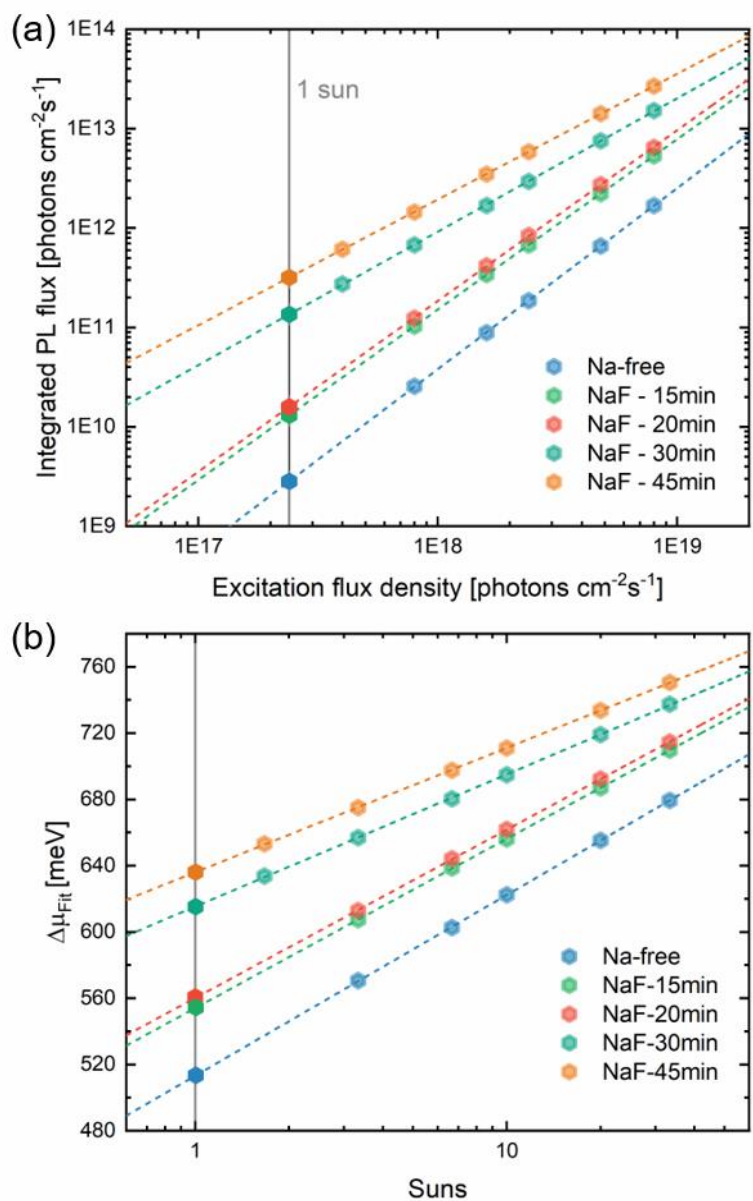

**Figure S6.** Extrapolation of PL flux and quasi-Fermi level splitting to 1sun for differently Na treated Cu(In,Ga)Se$_2$ single crystalline layers. The light colour dots are the measured values and the dark colour dots are the values extrapolated to 1 sun.

## S6. Time-resolved photoluminescence (TRPL)

In order to extract lifetime values, TRPL decays were fitted with a biexponential function after background subtraction. The time constants $\tau_1$ and $\tau_2$ obtained after the fittings are summarized in Table S3. At least two spots were measured for each sample and the values presented in the main text are the average. Only $\tau_2$ was considered as the effective bulk lifetime. An example of the fitting results can be seen in figure S7.

**Table S3**. Summary of the time constants $\tau_1$ and $\tau_2$ obtained after the fitting routine to the TRPL decays using a biexponential model.

| Sample | $\tau_1$ [ns] | $\tau_2$ [ns] | Average $\tau_1$ [ns] | Average $\tau_2$ [ns] |
| --- | --- | --- | --- | --- |
| As-grown | 0.58 | 1.83 | 0.61 | 1.81 |
| As-grown | 0.57 | 1.87 | | |
| As-grown | 0.68 | 1.73 | | |
| NaF – 15 min | 1.69 | 4.89 | 1.74 | 4.88 |
| NaF – 15 min | 1.79 | 4.88 | | |
| NaF – 20 min | 1.59 | 5.9 | 1.75 | 5.47 |
| NaF – 20 min | 1.71 | 5.02 | | |
| NaF – 20 min | 1.94 | 5.48 | | |
| NaF – 30 min | 1.66 | 5.24 | 1.67 | 5.43 |
| NaF – 30 min | 1.69 | 5.62 | | |
| NaF – 45 min | 1.44 | 6.55 | 1.46 | 6.42 |
| NaF – 45 min | 1.51 | 6.12 | | |
| NaF – 45 min | 1.44 | 6.6 | | |

$$I(t) = A_1 * \exp\left(-\frac{x}{\tau_1}\right) + A_1 * \exp\left(-\frac{x}{\tau_2}\right)$$

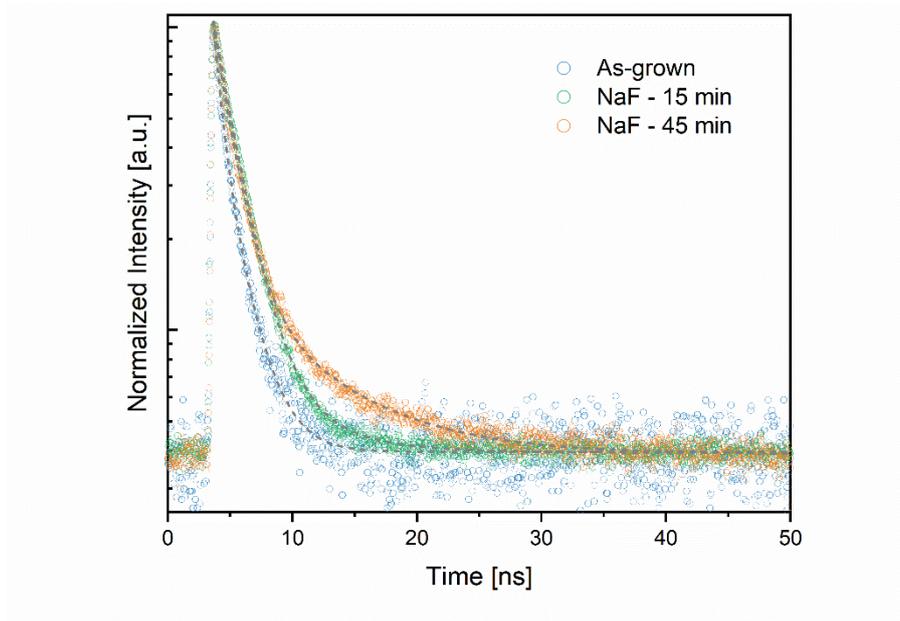

**Figure S7.** Time-resolved photoluminescence measurements of sample CIGSe273 with different sodium contents. Dashed lines represent best fits to the data using a biexponential model.